\documentclass[aps,twocolumn,groupedaddress,floatfix]{revtex4}

\usepackage{epsfig}

\begin{document}

\title{Real-time Monte-Carlo simulations for dissipative tight-binding
systems\break
 and time local master equations}

\author{Lothar M{\"u}hlbacher$^1$}
\author{Charlotte Escher$^1$}
\author{Joachim Ankerhold$^{1,2}$}
\affiliation{${}^1$Physikalisches Institut,
 Albert-Ludwigs-Universit{\"a}t,
 D-79104 Freiburg, Germany\\
 ${}^2$ Service de Physique de l'Etat Condens\'{e},
DSM/DRECAM, CEA Saclay, 91191 Gif-sur-Yvette, France}

\date{Date: \today}

\begin{abstract}
The numerically exact path integral Monte Carlo approach for the real-time
 evolution of dissipative quantum systems (PIMC), particularly suited for
 systems with discrete configuration space (tight-binding systems), is extended
 to treat spatially continuous and correlated many-body systems. This way, one has to
 consider generalized tight-binding lattices with either non-equidistant
 spacing or in higher dimensions, which in turn allows to analyze to what
 extent Markovian master equations can be applied beyond the usually studied
 spin-boson type of models.
\end{abstract}


\maketitle

\newpage

\section{Introduction}

Quantum Brownian motion is much more involved than its classical analog since
in general tractable equations of motion do not exist \cite{weiss}. Progress
can be made in two limiting ranges, namely, in the realm of weak dissipation
and in the opposite one of strong friction. In the former case a perturbative
treatment has led to a variety of Markovian weak-coupling master equations
\cite{qmoptics}, among them the famous Lindblad \cite{lindblad} and the
Redfield \cite{redfield} equations. Successful applications include quantum
optical systems, decoherence for solid-state based quantum bits and
nonadiabatic dynamics in molecular systems, to name but a few. Strong friction
has been explored only recently \cite{qsmolu1} with a growing amount of
research since then \cite{qsmolu2,qsmolu2a}. There, the quantum Smoluchowski
equation, a sort of Markovian master equation as well, allows to investigate
condensed phase dynamics at lower temperatures e.g.\ in soft matter and
mesoscopic systems \cite{qsmolu3}.

A formally exact description of open quantum systems valid
for all temperatures and damping strengths is provided by
the path integral approach initiated by the work of
Feynman and Vernon \cite{feynman} and developed in detail
in the 1980s \cite{caldeira,report,weiss}. The approach
has been utilized in numerous applications especially in
condensed phase systems, e.g.\ to reveal the
non-exponential decay of low temperature correlation
functions, and has further been exploited to consistently
derive the Markovian master equations mentioned above
\cite{qsmolu1,karrlein}. A real challenge, however, has
been to evaluate the formally exact expression for the
reduced density matrix in parameter regions where
analytical progress and perturbative simplifications are
prohibitive. With the increasing complexity of designed and
controllable quantum systems on the nanoscale, e.g.\ for
quantum information processing or molecular electronics,
the issue of efficient numerical procedures in the {\em
real-time domain} becomes a very crucial one. Indeed,
important achievements have been made in the last decade
with the development of advanced path integral Monte-Carlo
methods (PIMC) \cite{mc1}, the quasi-adiabatic propagator
scheme (QUAPI) \cite{quapi}, stochastic Schr\"odinger
equations \cite{strunz}, and basis set-methods
\cite{thoss}.

In this context a certain class of systems has been
extensively studied, namely, systems with discrete
configuration space, also coined tight-binding systems
(TBSs). For these systems quantum diffusion takes place on
a lattice, where the sites are coupled by tunneling
amplitudes, an important case being the restriction to
nearest neighbor coupling. The simplest example is the
well-known spin-boson model \cite{fisher} with
applications from condensed matter physics to electron
transfer reactions. The fundamental role of TBSs follows
from the diversity of realizations in physics and chemistry
\cite{weiss}. Transport properties in general multistable
systems at sufficiently low temperatures can be described
within TBS models leading to relations to the Kondo
problem and the Luttinger liquid model. Further, TBSs serve
as archetypical models to study quantum phase transitions
in correlated systems, as e.g.\ for Hubbard type of models.
Remarkably, even a large class of continuous systems can
be mapped exactly onto TBSs by means of duality
transformations \cite{weiss2}. Based on the path integral
representation exact non-Markovian master equations for
TBSs \cite{weiss} have been derived, which provide the
starting point for perturbative treatments such as the
non-interacting blip approximation (NIBA) and reduce to
Markovian ones even for moderate dissipation and low
temperatures \cite{weiss3}, i.e.\ far from the limiting
ranges addressed above.

The PIMC approach is particularly suited to treat the dissipative real-time
evolution of TBSs numerically exactly. While former applications were
restricted mainly to two and three state models \cite{egger}, recently, we
substantially improved the approach to apply to more complex systems
\cite{muehl1,muehl2}, such as single charge transfer across long
one-dimensional molecular chains including impurities and external driving
fields. Moreover, the simulation time range could be extended to cover
basically all relevant time scales of the dissipative dynamics. We have been
thus in a position to directly access the range of validity of Markovian master
equations for TBSs, which in turn are extremely helpful to reveal the relevant
physical processes behind the numerical data. The goal of this paper is now
twofold: On the one hand we push the PIMC procedure even further and present
first results for the dissipative dynamics of spatially continuous and of
correlated many-body systems; on the other hand, we give arguments to what
extent Markovian master equations can be used in these more involved
situations. We will see that this latter question will directly lead us to
consider generalized tight-binding lattices with non-equidistant spacing
between sites or in higher dimensions.

The article is organized as follows. We start in Sec.~\ref{sec1} with a brief
summary of the path integral representation for open quantum systems and
continue in Sec.~\ref{sec2} to collect the main ideas of the PIMC
scheme. Sec.~\ref{sec3} deals with known results for non-Markovian and
Markovian master equations for TBSs, the applicability of which for
one-dimensional chains is illustrated. Then, in Sec.~\ref{sec4} spatially
continuous systems are discussed, before we come to the correlated many-body
dynamics in Sec.~\ref{sec5}. At the end some conclusions are given.

\section{Dynamics of dissipative quantum
systems}\label{sec1}

The standard approach \cite{weiss} for the inclusion of dissipation into a
quantum mechanical formulation starts from a system+reservoir model
\begin{equation} \label{sm1}
H=H_S+H_R+H_I
\end{equation}
with a system part $H_S$, an environmental part $H_R$, and a system-bath
interaction $H_I$. The reservoir (heat bath) is mimicked by a quasi-continuum
of harmonic oscillators bilinearly coupled to the system:
\begin{equation} \label{hamilton}
 H_R + H_I
= \sum_\alpha \left[ {P_\alpha^2 \over 2m_\alpha}
 + {1\over2}m_\alpha\omega_\alpha^2 \left(X_\alpha+\frac{c_\alpha\, \hat{q}}{m_\alpha \omega_\alpha^2}
 \right)^2 \right] \; ,
\end{equation}
where $\hat{q}$ denotes a system operator corresponding to
a one-dimensional degree of freedom. Dissipation appears
when one considers the reduced dynamics by properly
eliminating the bath degrees of freedom, i.e.,
\begin{equation} \label{reduced}
\rho(t)={\rm Tr}_R\left\{{\rm e}^{-iHt/\hbar}\, W(0)\,
{\rm e}^{iHt/\hbar}\right\}
\end{equation}
with an initial density matrix $W(0)$ of the total system. It turns out that
for the reduced dynamics the environmental parameters enter only via the
spectral density
\begin{equation}
J(\omega) = \frac{a^2\, \pi}{2\hbar} \sum_\alpha
\frac{c_\alpha^2}
 {m_\alpha\omega_\alpha} \delta(\omega-\omega_\alpha) \;,
\end{equation}
which effectively becomes a continuous function of $\omega$ for a
condensed-phase environment. Note that in the above definition of the spectral
density we have included a factor $a^2/\hbar$ with $a$ being a proper length
scale which is convenient for the treatment of TBSs. The Gaussian statistics of
the isolated environment is determined by the complex-valued bath
autocorrelation function which for real time $t$ reads
\begin{eqnarray} \label{lz}
L(t) &=&\frac{a^2}{\hbar^2}\left\langle \left(\sum_\alpha
c_\alpha
X_\alpha(t)\right)\left(\sum_\alpha c_\alpha X_\alpha(0)\right) \right\rangle_\beta\nonumber\\
&=& {1\over\pi}\int_0^\infty \!d\omega\, J(\omega)
{\cosh[\omega(\hbar\beta/2-it)] \over
\sinh(\hbar\beta\omega/2)} \;,
\end{eqnarray}
where $\beta=1/k_B T$.

In the sequel we focus on systems evolving in a
discretized configuration space with respect to the
pointer variable $q$ and thus consider the population
$P(q_f,t)$ of a ``lattice site'' $q_f$ determined by the
diagonal part of the reduced density matrix, i.e.,
\begin{equation} \label{populations}
P(q_f,t) = {\rm Tr}\left\{ |q_f\rangle\!\langle q_f| \,
\rho(t)\right\} \;,
\end{equation}
which is normalized $\int dq P(q,t) = 1$. Accordingly, the
initial density matrix of the total compound $W(0)$ in
(\ref{reduced}) is taken to be
\begin{equation} \label{initial_perp_P}
W(0) = Z_R^{-1} |q_i\rangle\!\langle q_i|\, e^{-\beta
(H_R-q_i \mu{\cal E})} \;
\end{equation}
with the partition function of the isolated reservoir
$Z_R$. The bath is equilibrated according to a localized
initial state of the system on a lattice site $q_i$, where
$ \mu {\cal E} = \sum_\alpha c_\alpha X_\alpha$ so that
e.g.\ for electron transfer in a polar environment, $\mu$
is the electronic dipol moment and ${\cal E}$ the
collective dipol moment of the bath. Generalizations to
delocalized initial states for the system are
straightforward \cite{report}.

The path integral representation provides a formally exact expression for the
reduced dynamics and is thus the starting point for a numerically exact
Monte-Carlo (MC) scheme. Along the lines sketched above, the bath degrees of
freedom are eliminated exactly to arrive at the reduced dynamics. As shown in
Ref.~\cite{weiss}, one thus obtains for Eq.~(\ref{populations})
\begin{equation} \label{populations path-integral}
P_{s_f,s_i}(t) =  \oint \!{\cal
D}\tilde{s}\;\delta_{\tilde{s}(t),s_f}
  \exp\left\{ {i\over\hbar}S_S[\tilde{s}] -
 \Phi[\tilde{s}] \right\} \;.
\end{equation}
Here the path integration runs over closed paths $\tilde{s}(\tilde{t})$
connecting $\tilde{s}(0)=s_i$ with $\tilde{s}(t)=s_f$ along the real-time
contour $\tilde{t} \in 0 \rightarrow t \rightarrow 0$, which combines the
forward and backward paths $s(t')$ and $s'(t')$, respectively. Furthermore,
$S_S[s]$ denotes the action of the free system. The influence of the traced-out
bath is completely encoded in the {\sl Feynman-Vernon influence functional}
$\Phi[s]$ \cite{feynman}
\begin{eqnarray} \label{influence-exponent}
\Phi[q, q'] &=& \int_0^{t}\!dt'\! \int_0^{t'}\!dt'' [q(t')
- q'(t')]
\ [{L}(t'-t'')q(t'') \nonumber\\
&& \qquad {}- {L}^\ast(t'-t'')q'(t'')] \nonumber\\
&& {}+ i{{\hat{\mu}} \over 2}\int_0^{t}\!dt' [q^2(t') -
q'^2(t')] \;,
\end{eqnarray}
where
\begin{equation} \label{muu}
{\hat{\mu}} = {2  \over \pi} \int_0^\infty\!d\omega
{J(\omega) \over \omega}  \;.
\end{equation}
The influence functional introduces long-ranged nonlocal
interactions among the system paths so that in general an
explicit evaluation of the remaining path integral in
Eq.~(\ref{populations path-integral}) is possible only
numerically. In this situation the PIMC method has been
proven as a very promising approach to obtain numerically
exact results even in regions of parameter space where
other approximate methods fail.

\section{PIMC simulation method}
\label{sec2}

A prerequisite for an efficient numerical algorithm is an
appropriate discretization of time and configuration
space. The latter one is intrinsically given for
multistable systems in the tight binding limit, where the
relevant states are strongly localized in position, only
very weakly coupled by tunneling, and energetically well
separated from the rest of the spectrum.  For spatially
continuous systems supporting delocalized states the
situation is less obvious, but in case of a discrete
energy spectrum a mapping onto a generalized tight-binding
lattice applies as well for lower temperatures
\cite{quapi,grifoni}. In any case, the configuration space
variable can then be written as $q(t)=a \cdot s(t)$ with a
typical length scale $a$ and a dimensionless variable
$s(t)\in\{q_1, \ldots, q_d\}$ with $-S= q_1<q_2<\cdots
<q_d=S$ according to a $d$-level system ($d$LS). Hence,
the system Hamiltonian reads
\begin{equation}
H_{d{\rm LS}}=\hbar E_z - \hbar S_x \;,
\end{equation}
where $E_z$ describes the energetic distribution of the sites according to
$E_z|q_\mu\rangle= \epsilon_\mu |q_\mu\rangle$ and $S_x$ the couplings between
them $\Delta_{\mu\nu} = \langle q_\mu|S_x|q_\nu\rangle, \mu\neq \nu$. In
particular, in case of $q_{\mu+1}-q_\mu=1$ and nearest neighbor coupling only
one recovers a $(2S+1)$-spin-boson model.

For the discretization in time, the time axis is sliced via $r$ uniformly
spaced points with discretization steps $\tau = t/r$. The path integral in
Eq.~(\ref{populations path-integral}) then becomes
\begin{equation} \label{pop-discretized}
P_{s_f, s_i}(t) = \sum_{\{s_j\}} \delta_{s_{r+1},s_f}
\rho[\{s_j\}]
\end{equation}
with
\begin{equation} \label{density}
\rho[\{s_j\}] = \left[\prod_{k=1}^{2 r}
K(s_k,s_{k+1})\right]e^{-\Phi[\{s_j\}]} \;.
\end{equation}
The sum runs over all realizations of the discretized spin path $\{s_j\} =
\{s_1\equiv s_i,s_2,\dots,s_{2r}, s_{2r+1}\equiv s_i\}$, and $K(s_j,s_{j+1})$
denotes the coordinate representation of the free $d$LS propagation over the
time interval $\tau$, i.e.,
\begin{equation}
K(s,s',\tau)=\langle s|\exp(-i\tau H_{d\rm
LS}/\hbar)|s'\rangle \;.
\end{equation}
This propagator of the $d$LS Hamiltonian can be obtained from the eigenstates
\begin{equation} \label{syseigen}
H_{d\rm LS} |\phi_\alpha\rangle =
E_\alpha|\phi_\alpha\rangle\,, \quad \alpha=1,\ldots, d
\end{equation}
as
\begin{equation} \label{sysprop2}
K(s,s',\tau) = \sum_{\alpha=1}^d \langle
s|\phi_\alpha\rangle \langle\phi_\alpha|s'\rangle\ {\rm
e}^{-i \tau E_\alpha/\hbar} \;,
\end{equation}
which can be easily computed numerically once the $d$LS's parameters are
specified.

To arrive at a discretized form (in time) of the influence functional
(\ref{influence-exponent}), the sum and difference coordinates
\begin{equation} \label{sum and difference coordinates}
\eta(t') \equiv s(t') + s'(t') \;, \qquad \xi(t) \equiv
s(t') - s'(t')
\end{equation}
are introduced, which read $\eta(t') = \eta_j$ ($\xi(t') = \xi_j$) for $t' \in
[(j-1)\tau - \tau/2, (j-1)\tau + \tau/2]$ in their discretized form. The sum
paths are also considered as ``quasi-classical'', while the difference paths
capture quantum fluctuations \cite{weiss}. Equation~(\ref{influence-exponent})
finally can be written as
\begin{eqnarray} \label{influence}
\Phi[s_i, \eta, \xi] &=&
i\sum_{j=2}^r \xi_j \hat{X}^{(s_i)}_j\\
&&{}+ \sum_{j \ge k=2}^r \xi_j (i X_{j-k} \eta_k +
\Lambda_{j-k} \xi_k) \nonumber
\end{eqnarray}
where the kernels $\hat{X}^{(s_i)}_j$, $X_{j-k}$, and $\Lambda_{j-k}$ follow
from discretizing the twice integrated bath autocorrelation function $Q(t)$
defined by $\ddot{Q}(t) = L(t)$, $Q(0) = 0$ with $\dot{Q}(0) =
i\hat{\mu}/2$. In the sequel we consider a spectral density of the form
\begin{equation} \label{spectral}
J(\omega)= 2 \pi \alpha \omega {\rm e}^{-\omega/\omega_c}
\;,
\end{equation}
which is equivalent to ohmic damping with a cut-off frequency $\omega_c$. In
this case $Q(t)$ can be calculated analytically and one obtains
\begin{equation}
Q(t) = 2\alpha\Bigg[ \ln(1+i\omega_c t)
 - \ln\frac{\Gamma(\Omega+i t/\hbar\beta)\Gamma(\Omega-it/\hbar\beta)}
 {\Gamma^2(\Omega)} \Bigg]
\end{equation}
with $\Omega = 1 + 1/(\hbar\beta\omega_c)$ and the Gamma function $\Gamma(z)$.

Equations~(\ref{pop-discretized}), (\ref{density}) and (\ref{influence})
constitute a discretized form of the populations (\ref{populations}) and thus
provide a starting point for PIMC simulations. As it is well-known this method
is handicapped by the dynamical sign problem \cite{sign-problem}. It originates
from quantum interferences between different system paths $\{s_j\}$, causing a
small signal-to-noise ratio of the stochastic averaging procedure.

One approach to deal with this problem is based on the observation that the
quasi-classical paths $\{\eta_j\}$ in Eq.~(\ref{influence}) can be integrated
out as a series of $r-1$ matrix multiplications \cite{egger3}. This reduces the
degrees of freedom from the $2r-1$ variables $\{\eta_{2\le j \le r+1},
\xi_{2\le j \le r}\}$ to the $r-1$ quantum variables $\{\xi_{2\le j \le r}\}$
and therefore significantly improves the numerical stability of the
corresponding MC simulations. While this technique, which in fact is yet
another example of a blocking approach \cite{blocking}, works greatly for
dissipative two- and three-state systems \cite{egger}, however, the increasing
size of the corresponding matrices with the number of electronic sites turns
these multiplications into a quite time consuming task. Since they have to be
performed for every single update of the MC trajectory, the investigation of
larger systems again requires exceptionally long CPU times.

Nevertheless, this severe computational bottleneck can be profoundly alleviated
on physical grounds \cite{muehl1,muehl2}. Upon closer inspection one finds that
the possibility of rewriting the integration over the quasi-classical
coordinates in terms of simple matrix multiplications is due to the fact that
the real-valued part of the bath autocorrelation function (\ref{lz}) governs
only the {\em quantum coordinates} but not the quasi-classical ones. This
real-valued part, which eventually leads to a damping out of quantum
coherences, introduces a non-local self-coupling among the quantum coordinates
and is thus directly related to retardation effects, a main complication for
treating dissipative quantum systems. Accordingly, the retardation effects
influence the evolution of the quasi-classical coordinates contained in the
imaginary part of the influence functional to a significantly weaker extend
than that of the quantum coordinates. Neglecting them while generating the MC
trajectories leads to an only minor impairment of the sampling statistics, but
causes an almost complete decoupling between quantum and quasi-classical
coordinates. This in turn allows for an enormous speed-up of the matrix
multiplications. Accordingly, the sampling process could so far be accelerated
by a factor of approximately 100 with respect to the original approach
\cite{egger} (for further details, we refer to Refs.~\cite{muehl1,muehl2}),
thus opening the door to treat the reduced dynamics of much larger and even
many-body systems over sufficiently long times.

\section{Time-local master equations}\label{sec3}

From the exact expression (\ref{populations path-integral}) for the reduced
density simplifications can be derived in certain limits in terms of time-local
master equations. In the context discussed here, these are of crucial
importance, mainly since (i) master equations allow to study ranges in
parameter space where the Monte Carlo sampling is rather expensive, e.g.\ for
very weak or very strong friction, and (ii) they provide a basis to access the
physical processes underlying the numerical data also by means of analytical
techniques.

In the weak friction limit one imposes that the level broadening due to
friction is much smaller than $k_{\rm B} T$ {\em and} the typical level
separation. It is thus natural to work in the basis spanned by the energy
eigenstates of $H_S$ and to treat the coupling $H_I$ perturbatively. With
increasing coupling, however, the environment drives the system to its pointer
basis in which the system-bath coupling operator $q$ is diagonal. As we have
seen above in Sec.~\ref{sec1}, in this basis the path integral formulation
allows for a non-perturbative elimination of the reservoir degrees of
freedom. In case that friction is very strong and thus level broadening much
larger than level spacing and temperature, a master equation complementary to
the weak friction range, the quantum Smoluchowski equation, is again available
\cite{qsmolu1,lehle}.

What's about the intermediate regime? For TBSs discussed here, substantial
findings have been gained in the past decade \cite{weiss}. Namely, it was shown
that an {\em exact} retarded master equation exists, i.e.,
\begin{equation} \label{exactME}
\frac{dP_{\mu_f,\mu_i}(t)}{dt}=\sum_{\nu=1}^{d} \int_0^t
dt'\ \hat{\Gamma}_{\mu_f\nu}(t-t')\, P_{\nu,\mu_i}(t')\, ,
\end{equation}
where the kernels obey
$\hat{\Gamma}_{\mu\mu}=-\sum_{\nu\neq\mu}\hat{\Gamma}_{\mu\nu}$.
Basically they represent a power series in the couplings
$\Delta_{\mu \nu}$ with corresponding spin-path integrals.
To make the latter tractable, one applies the NIBA
\cite{fisher} or its generalization, the non-interacting
cluster approximation (NICA) \cite{weiss}, which has been
shown to be accurate in a variety of parameter ranges. For
instance, in case of an ohmic spectral density with a high
frequency cut-off, it captures quantum coherence as well
as pure relaxation dynamics. The corresponding kernels
$\tilde{\Gamma}_{\mu\nu}$ read in lowest order in the
intersite couplings
\begin{eqnarray} \label{grrate}
\tilde{\Gamma}_{\mu\nu}^{(2)}(t)&=&2 \Delta_{\mu\nu}^2\
\exp\left[-(q_\mu-q_\nu)^2\, Q'(t)\right]\nonumber\\ &&
\cos\left[(\epsilon_\mu-\epsilon_\nu) t+(q_\mu-q_\nu)^2\,
Q''(t)\right]\,
\end{eqnarray}
with $Q'={\rm Re}Q(t)$ and $Q''={\rm Im}Q(t)$. The practical use of
(\ref{exactME}) is limited though due to the retardation.

Now, for sufficiently strong dissipation and fast enough bath modes the kernel
falls off on a time scale much shorter than the time scale on which the
relevant reduced dynamics occurs so that we may set in (\ref{exactME})
$P_{\nu,\mu_i}(t')\approx P_{\nu,\mu_i}(t) $ and
\begin{equation} \label{grfull}
\Gamma_{\mu\nu}=\int_0^\infty dt\,
\tilde{\Gamma}_{\mu\nu}(t)\, .
\end{equation}
Accordingly, (\ref{exactME}) reduces to a simple Markovian rate equation
\begin{equation} \label{differential rate equation}
\dot{{\mathbf P}}(t) = A\, {\mathbf P}(t) \;,
\end{equation}
with a rate matrix $A$ consisting of the individual rates
$\Gamma_{\mu\nu}$. For nearest neighbor coupling, the golden rule rates
$\Gamma_{\mu\nu}^{(2)}$ describe a sequential hopping process, while all higher
order contributions to $\Gamma_{\mu\nu}$ capture long-range hopping termed
superexchange. Note that in contrast to weak coupling master equations as e.g.\
the Redfield equations, (\ref{exactME}) and (\ref{differential rate equation})
together with the corresponding transition rates apply also to the range of
moderate to strong bath coupling and very low temperatures.

\begin{figure}
\epsfig{file=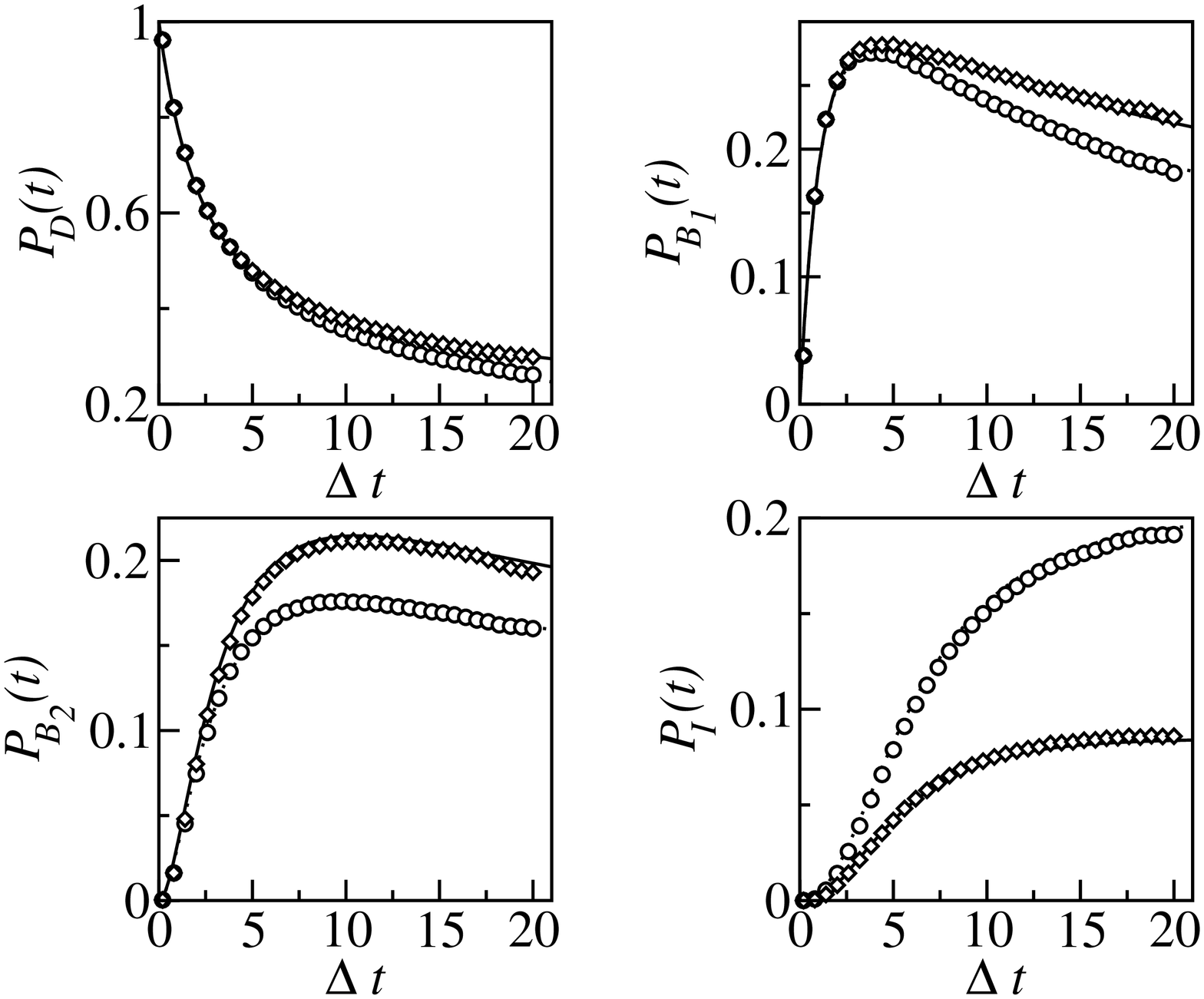,angle=-90, width=10cm}
\epsfig{file=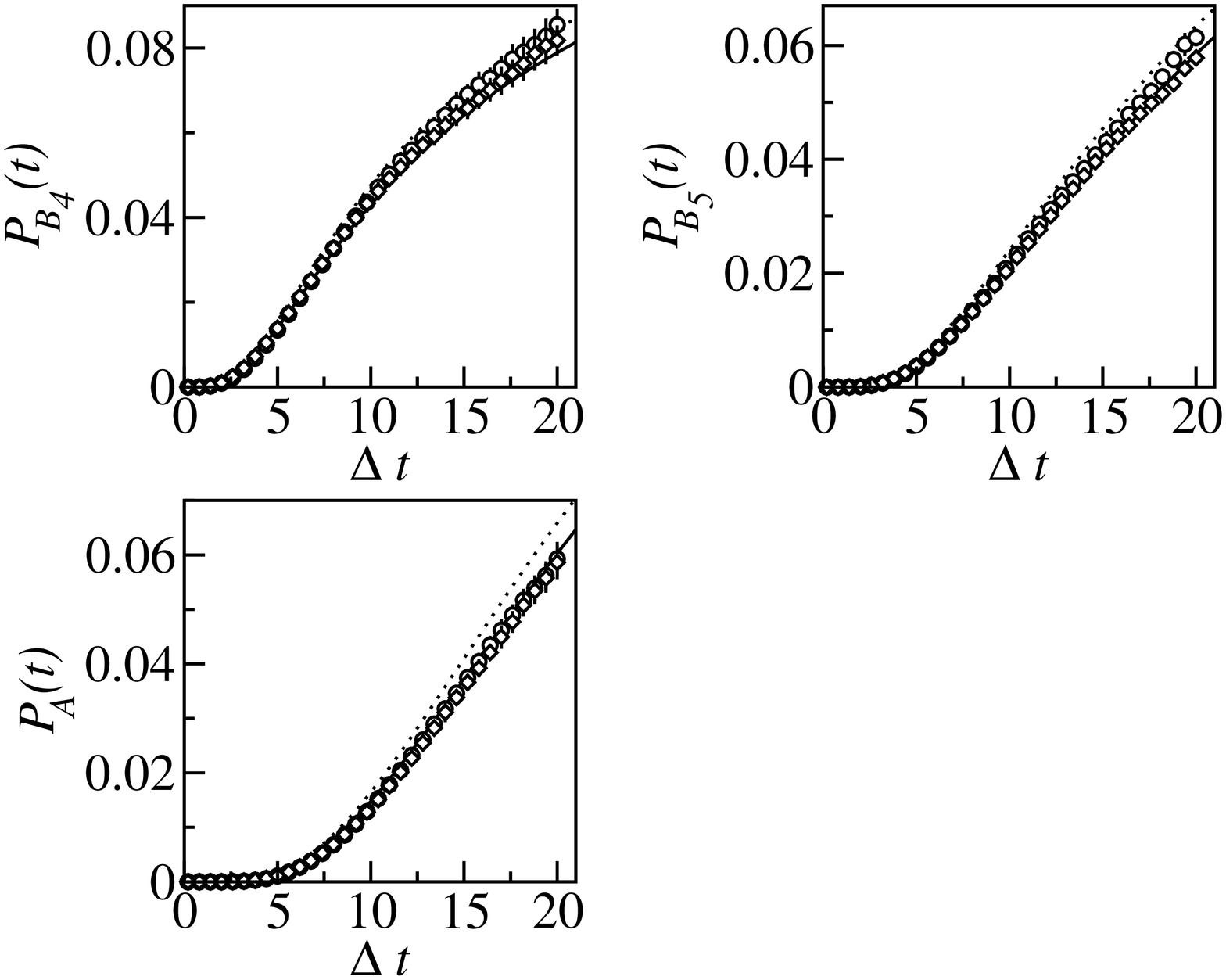,angle=-90,width=10cm}
\caption[]{\label{fig1} Populations along a molecular
chain with $d=7$ sites obtained from PIMC simulations
(symbols) and from the local master equation approach
(lines). Donor (D) and acceptor (A) are connected by a
bridge with an impurity (I) at its center. The impurity
has an onsite energy relative to D/A of
$\epsilon_I/\Delta=+5$ (PIMC: diamonds; master equation:
solid lines) and $\epsilon_I/\Delta=-5$ (PIMC: circles;
master equation: diamonds). Other parameters are
$\alpha=0.1, \Delta\hbar\beta=0.1, \omega_c/\Delta=5$, see
text for details.}
\end{figure}

As an explicit example of (\ref{differential rate
equation}) we consider a one-dimensional tight-binding
lattice with $d=7$ sites, spacing 1, and constant nearest
neighbor coupling corresponding to a spin-3-boson system
(for details see \cite{muehl2}). This model can be seen as
a simple realization of electron transfer along a
molecular chain subject to a dissipative environment
\cite{jortner,muehl1}. Initially, the charge is localized
at the donor site $s_i=q_1=-3$ and $P_{s_f,-S}(t)$ monitors
the dynamics towards the acceptor at $q_7=3$. Specifically,
donor and acceptor are linked by a bridge with an impurity
at its center. Donor/acceptor have vanishing onsite
energy, the bridge sites are elevated by
$\epsilon_B/\Delta=2.5$, and the impurity site has
$\epsilon_I/\Delta=\pm 5$. As seen in fig.~\ref{fig1} the
exact PIMC data are very accurately described by the
time-local master equation (\ref{differential rate
equation}) despite the fact that all parameters are chosen
such that one is close to a coherent/incoherent transition
(for the given values of $\omega_c$ and $\alpha$
coherences appear at inverse temperatures
$\hbar\beta\Delta\approx 0.3$ and larger) and
$\omega_c/\Delta=5$ is far from the scaling limit.

\section{Dynamics of spatially continuous systems}\label{sec4}

In order to go beyond the spin-boson type of models
(one-dimensional tight binding lattice, equidistant
spacing, constant nearest coupling), we start by
presenting an approach to capture the dynamics of
spatially continuous systems within the PIMC procedure
outlined above. It turns out that this method applies to
systems with a discrete energy spectrum. The basic idea is
simple: For sufficiently low temperatures only the lowest
lying states of the isolated system can be assumed to take
part in the dynamics; thus, the full Hilbert space $H_S$
can effectively be truncated to a subspace $H_S^{(N)}$
spanned by the $N$ lowest lying eigenstates. One then has
to find a proper basis in this subspace, obtained by a
unitary transformation from the eigenstate basis, in which
a stochastic sampling of the path integrals
(\ref{populations path-integral}) can be performed.

This sort of reduction is not new. In fact, the spin-boson
model can be seen as originating from a double well
potential, where only the two lowest lying states are
taken into account. Of course, the goal here is to go
beyond: We want to capture the dissipative dynamics from
very low up to moderate temperatures. For very high
temperatures semi-classical or classical methods apply
anyway. Further, we are interested in the regime of
moderate friction, where neither of the known approximate
formulations work so that a numerical treatment has to
start from the exact reduced dynamics (\ref{populations
path-integral}). The proper basis for the sampling in the
restricted Hilbert space is then the basis that
diagonalizes the operator $\hat{q}$ in $H_S^{(N)}$. This
idea to treat spatially continuous systems has been first
applied in the numerical QUAPI approach \cite{quapi} and
has been recently analyzed in \cite{grifoni} to derive a
generalized master equation of the type given above
(\ref{exactME}). Hence, we sketch here only the central
results briefly and refer to these previous works for
further details.

In the $N$-dimensional subspace $H_S^{(N)}$ the pointer
basis is defined as
\begin{equation}
\hat{q}|q_\nu\rangle=a q_\nu\, |q_\nu\rangle\ ,\
\nu=1,\ldots, N \label{pointer}
\end{equation}
and obtained from the eigenstate basis $\{|n\rangle\}$ by
diagonalizing a matrix with entries $\langle
n|\hat{q}|m\rangle$, $n,m=1, \ldots, N$. Its eigenvalues
provide the dimensionless $q_\nu$, while the eigenvectors
are given as $|q_\nu\rangle=\sum b_{\nu n}|n\rangle$. The
basis set $\{|q_\nu\rangle\}$ is also known as the
DVR-basis (Discrete Value Representation). By representing
the Hamiltonian in this DVR-basis, one obtains ``onsite
energies'' $\hbar\epsilon_\nu=\langle
q_\nu|H_S|q_\nu\rangle$ and ``intersite-couplings''
$\Delta_{\nu\mu}=\langle q_\nu|H_S|q_\mu\rangle/\hbar$. The
original system is thus mapped onto a generalized
$N$-dimensional tight-binding lattice with {\em
non-equidistant} sites at $q_\nu, \nu=1, \ldots, N$ and
{\em non-nearest neighbor} couplings $\Delta_{\nu\mu}$.
Now, to implement this TBS into the PIMC algorithm, one
has to take into account the non-equidistant lattice by
re-defining the $\xi_j$ and $\eta_j$ variables properly.

A rough estimate for the validity of the truncation procedure to the lowest $N$
eigenstates of the isolated system is given by the conditions that (i) the
level broadening due to friction is of the same order as the level spacing or
smaller and that (ii) the temperature is sufficiently low
$N\hbar\beta\omega_0\gg 1$. While the second condition is obvious, the first
one originates from the fact that for very strong friction the system tends to
the classical limit again, see \cite{qsmolu1}.

By way of example, we consider in the sequel a harmonic
oscillator with mass $M$ and frequency $\omega_0$, so that
the typical length scale is $a\equiv
q_0=\sqrt{\hbar/M\omega_0}$. While a detailed account of
the implementation into the PIMC algorithm and results of
simulations for specific observables, particularly in
comparison with analytical results, will be given
elsewhere, here, our main interest is to explicitly show
that apart from the conditions for the applicability of
time local master equations known for spin-boson models,
the treatment of continuous systems via the described
mapping comes with an additional complication. Formally,
this issue has also been addressed recently in
\cite{grifoni}. Here, we directly compare numerically
exact PIMC data with results from the simplified master
equation. For this purpose the initial state is chosen as
one of the DVR-states which suffices to study the time
evolution of the populations on the sites $q_\nu$.

\begin{figure}
\epsfig{file=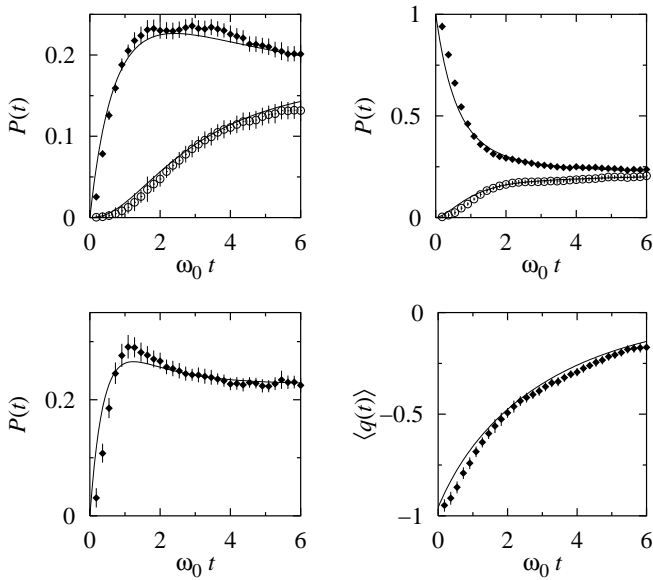, width=9cm}
\caption[]{\label{fig2} DVR-site populations for a
harmonic oscillator with $N=5$ eigenstates obtained from
PIMC simulations (symbols) and a master equation approach
(solid lines). Top left: $P_1$ (diamonds) and $P_5$
(circles), top right: $P_2$ (diamonds) and $P_4$
(circles), bottom left: $P_3$, bottom right: average
position, see text for details. }
\end{figure}

\begin{figure}
\epsfig{file=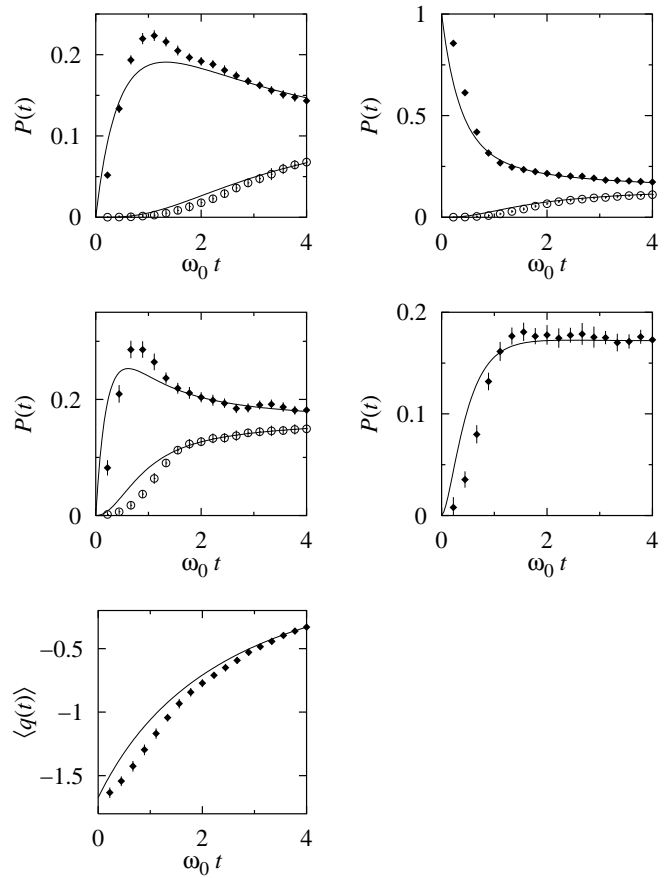, width=9cm} \caption[]{\label{fig3}
Same as in fig.~(\ref{fig2}) but for $N=7$. Top left:
$P_1$ (diamonds) and $P_7$ (circles), top right: $P_2$
(diamonds) and $P_6$ (circles), middle left: $P_3$
(diamonds) and $P_5$ (circles), middle left: $P_4$,
bottom: average position, see text for details. }
\end{figure}

Results for $N=5$ and $N=7$ are depicted in
figs.~\ref{fig2}, \ref{fig3} for a bath with
$\omega_c/\omega_0=5$, $\alpha=0.2$, and
$\hbar\beta\omega_0=0.2$. For this parameter set we have
shown recently \cite{muehl1,muehl2} that time local master
equations capture the exact dynamics of linear chains
rather accurate even though one is close to an
incoherent/coherent transition for the reduced dynamics.
The same is true here for $N=5$, where we used
(\ref{differential rate equation}) together with the
hopping rates (\ref{grfull}). In contrast, for $N=7$
deviations appear in the short to intermediate time
regime. For times of the order of $1/\omega_c$ and shorter
($\omega_0 t\leq 0.2$), these are caused by adiabatic
effects in the bath well-known from the dynamics in
ordinary tight binding lattices. In the intermediate time
range, however, where for certain populations pronounced
maxima occur, deviations must be ascribed a specific
property related to the mapping onto a tight binding
lattice with non-equidistant spacing. As mentioned above,
this mapping requires a re-definition of the $\xi$ and
$\eta$ variables, originally defined for a tight binding
lattice with equidistant spacing of length $a$.

Qualitatively, the situation is the following: Since the variance of the $N$th
eigenstate is roughly $(\Delta q^2)_N \approx N q_0^2$, the box in position
space covered by $N$ states has a typical width $L\approx 2\sqrt{N} q_0$. The
mean spacing between adjacent lattice sites is thus $a_{\rm eff} \approx L/N=2
q_0/\sqrt{N}$. Hence, compared to a lattice with equidistant spacing of length
$a=q_0$, the re-definition of the $\xi$ and $\eta$ variables in the influence
functional (\ref{influence}) comes with an additional factor of the order
$(2/\sqrt{N})^2$, which basically renormalizes the damping
kernel. Consequently, for fixed bath parameters and increasing $N$, the
effective coupling constant between bath and discrete system is not constant,
but decreases as $4\alpha/N$. The system is thus driven into the range of weak
coupling to the bath \cite{grifoni}, where retardation effects due to quantum
coherences become more substantial. Hence, the limit to a spatially continuous
system $N\to \infty$ is non-trivial and must be performed by keeping $\alpha/N$
constant. Specifically, by comparing the spectral density (\ref{spectral}) with
the one usually introduced for spatially continuous systems \cite{weiss}, one
finds
\begin{equation}
\frac{4\alpha \pi}{{N}} \approx\frac{\gamma}{2\omega_0}\ ,
\label{ohmic}
\end{equation}
where $\gamma$ denotes the macroscopic damping constant
appearing in the classical Langevin equation. Note that
the above relation cannot be seen as a strict equality
since the spacing between adjacent DVR-sites varies
slightly and is typically smaller deep inside the
potential well. The value $\alpha=0.2$ chosen above
corresponds for $N=5$ to $\gamma/\omega_0\approx 1$ and
for $N=7$ to $\gamma/\omega_0\approx 0.7$. Now, a rough
estimate for the validity of time local master equations
can be gained by assuming that the exponential in
(\ref{grrate}) must fall off on a time scale sufficiently
shorter than that of the dynamics of $P_\mu(t)$. For
moderate and low temperatures, this leads to
$1/(\omega_c\sqrt{\alpha/N}){\textstyle {\lower 2pt
\hbox{$<$} \atop \raise 1pt \hbox{$ \sim$}}} 1/\omega_0$
and thus $N {\textstyle {\lower 2pt \hbox{$<$} \atop
\raise 1pt \hbox{$ \sim$}}} \alpha (\omega_c/\omega_0)^2$.
In accordance with this relation, the PIMC data presented
in fig.~\ref{fig2}, \ref{fig3} for $N=5$ can still be
captured quantitatively by (\ref{differential rate
equation}), while for $N=7$ only a qualitative agreement
can be seen. To perform the continuum limit is thus not an
easy task and certainly deserves further research.

\section{Correlated two-particle dynamics }\label{sec5}

The dissipative real-time evolution of interacting many body systems has been
left basically untouched so far. In certain limits, e.g.\ for very strong
repulsive interactions, results could be derived from a simple hopping model
\cite{petrov}, but the intimate interplay between direct interaction,
interaction mediated by the environment and dissipation has not been
accessible. Here, for the first time we present PIMC results for the dynamics
of two interacting particles along the lines described above. As in the
previous section, in the sequel the formulation is outlined only briefly and we
focus on conceptual properties related to time-local master equations. Further,
we consider the case of indistinguishable particles, called charges henceforth,
so that the quantum nature of the particle statistics matters. Physically, the
situation refers to interacting spinless fermions or interacting bosons.

The free system is taken as a tight-binding lattice with
spacing 1 and nearest neighbor coupling, where the two
charges can be placed on $d=2S+1$ sites. Accordingly, the
two-charge Hamilton operator reads
\begin{equation}
H_{d{\rm LS}}^{(2)}=\hbar\left\{{ E}_z - [{ S}_x^{(a)}+{
S}_x^{(b)}]+{ U}\right\} \label{2hamilton}
\end{equation}
where ${ E}_z$ and ${ S}_x^{(j)}, j=a, b$ are straightforward generalizations
of the corresponding operators introduced above for the single particle case,
while ${ U}$ describes a site dependent Coulomb interaction specified
below. The coupling to the bath is determined by the total dipol-moment of the
system and thus follows directly from (\ref{hamilton}):
\begin{eqnarray}
H&=& H_{d{\rm LS}}^{(2)}+\sum_\alpha\left\{{P_\alpha^2
\over 2m_\alpha}+
 \frac{1}{2 m_\alpha\omega_\alpha^2}\right.\nonumber\\
&&\hspace{1cm}\left.\times \left[X_\alpha-
 \frac{c_\alpha a}{m_\alpha\omega_\alpha^2}
 \left({ S}_z^{(a)}+{ S}_z^{(b)}\right)
 \right]^2\right\}
 \label{total2}
 \end{eqnarray}
with ${ S}_z^{(j)}|s^{(j)}\rangle=s^{(j)}\, |s^{(j)}\rangle, j=a, b$.

Since the two charges are indistinguishable it is convenient for the PIMC
simulation to work not in the single particle product basis, but rather in the
basis of the many body states. For this purpose we introduce
\begin{equation}
\{|s^{(a)}\rangle|s^{(b)}\rangle\} \longrightarrow \{|s,
\hat{s} \rangle\} \label{2basis}
\end{equation}
with $s\leq \hat{s}$. The advantage of this representation
is that the $d (d+1)/2$ states $\{|s,\hat{s}\rangle\}$ are
orthogonal in contrast to the original ones. Accordingly,
$H_{d{\rm LS}}^{(2)}$ takes the form
\begin{equation} \label{dhamilton}
H_{d{\rm LS}}^{(2)} = \hbar\left[{E}_z - { S}_x + { U}
\right]
\end{equation}
where now
\begin{eqnarray}
{ E}_z |s,\hat{s}\rangle
&=& (\epsilon_s + \epsilon_{\hat{s}})|s,\hat{s}\rangle \nonumber\\
{S}_x |s,\hat{s}\rangle &=&
\delta_{s=\hat{s}}(\Delta_{s-1}|s-1,s\rangle +
\Delta_s|s,s+1\rangle)
\nonumber\\
&& {} + \delta_{s \neq
\hat{s}}(\Delta_{s-1}|s-1,\hat{s}\rangle +
\Delta_{\hat{s}}|s,\hat{s}+1\rangle \nonumber\\
&& {}+\Delta_{s}|s+1,\hat{s}\rangle
 + \Delta_{\hat{s}-1}|s,\hat{s}-1\rangle) \nonumber\\
{ U} |s,\hat{s}\rangle &=&
\frac{1}{2}(u_{s,\hat{s}}+u_{\hat{s},s})|s,\hat{s}\rangle\,
. \label{new2h}
\end{eqnarray}
For an onsite Coulomb energy this model is identical to a
dissipative Hubbard model with two charges. Note, however,
that for the PIMC simulations a non-local interaction can
easily be taken into account.

Now, as seen above, the free system dynamics enters in the
discretized path integral formulation only through its
short time propagator which is conveniently represented in
the energy eigenbasis of $H_{d{\rm LS}}^{(2)}$ in
(\ref{dhamilton}), see (\ref{sysprop2}). Further, since in
(\ref{total2}) the two charges interact with the bath only
via the sum $S_z^{(a)}+S_z^{(b)}$ the corresponding change
to the many body basis in the influence functional is
straightforward. It amounts to introduce discretized sum
and difference paths according to
\begin{eqnarray}
\eta_j \equiv s_j + s'_j \;,
  && \xi_j \equiv s_j - s'_j \;, \nonumber\\
\hat{\eta}_j \equiv \hat{s}_j + \hat{s}'_j \;,
  && \hat{\xi}_j \equiv \hat{s}_j - \hat{s}'_j\,
\end{eqnarray}
such that the influence functional depends only on $\eta(t')+\hat{\eta}(t')$
and $\xi(t')+\hat{\xi}(t')$.

\begin{figure}
\epsfxsize=6cm
\epsffile{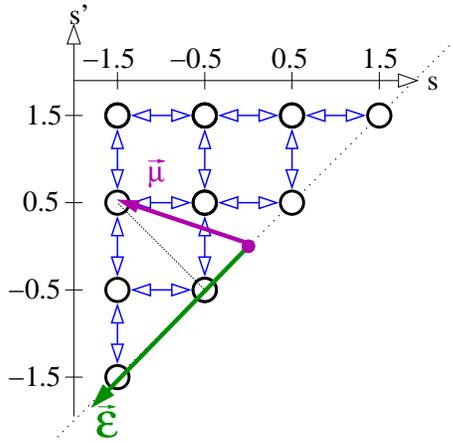} \caption[]{\label{fig4} Mapping of a
two charge system moving in a one-dimensional tight binding
lattice onto a one particle system diffusing on a
two-dimensional triangular lattice.}
\end{figure}

In principle, we could now start to implement the above
representation into a PIMC algorithm. However, before
doing so we go one step further and exploit the following
crucial property: {\em The dissipative dynamics of two
indistinguishable charges on a one-dimensional lattice
with $d$ sites is equivalent to the dissipative dynamics
of a single particle on a two-dimensional triangular
lattice with $d (d+1)/2$ sites}. To see this, one realizes
that the coupling to the bath $({ S}_z^{(a)}+{ S}_z^{(b)})
\, \mu{\cal E}$ with $\mu{\cal E}$ as in
(\ref{initial_perp_P}) can be written as
$\vec{\mu}\cdot\vec{\cal E}$ with the vector operators
$\vec{\mu}= (\mu\,{ S}_z^{(a)},\mu\,{ S}_z^{(b)})$ and
$\vec{\cal E }=({\cal E},{\cal E})$, thus being identical
to the system-bath coupling of a single particle on a
surface. Eventually, by a proper re-labeling of the
populations $P(s_i,s_f,\hat{s}_i,\hat{s_f};t)\to
\tilde{P}(l_i,l_f;t); \ l_i, l_f=1,\ldots, d(d+1)/2$ one
formally obtains the desired mapping. This is illustrated
in fig.~\ref{fig4} where the actual coupling with the bath
at a certain site is given by the projection of the
corresponding ``spin-vector'' $\vec{\mu}= (\mu s, \mu
\hat{s})$ onto $\vec{\cal E}$ as just described and each
site carries an onsite energy depending on $\epsilon_s +
\epsilon_{\hat{s}}+(u_{s,\hat{s}}+u_{\hat{s},s})/2$. Due
to the nearest neighbor-coupling, transitions can only
occur in the vertical and the horizontal direction,
respectively. Note that this mapping can be generalized to
fermionic systems as well.

To summarize, the advantages of the many body basis and the subsequent mapping
onto a single particle 2$d$-lattice are: (i) the PIMC algorithm developed for
single particles can be simply adapted to the case of two particles, (ii) the
configuration space to be sampled is reduced from $d^2$ to $d (d+1)/2$, and
(iii) the master equations introduced in the previous section can be directly
applied.

\begin{figure}
\epsfig{file=figure5a.ps,
width=8cm} \epsfig{file=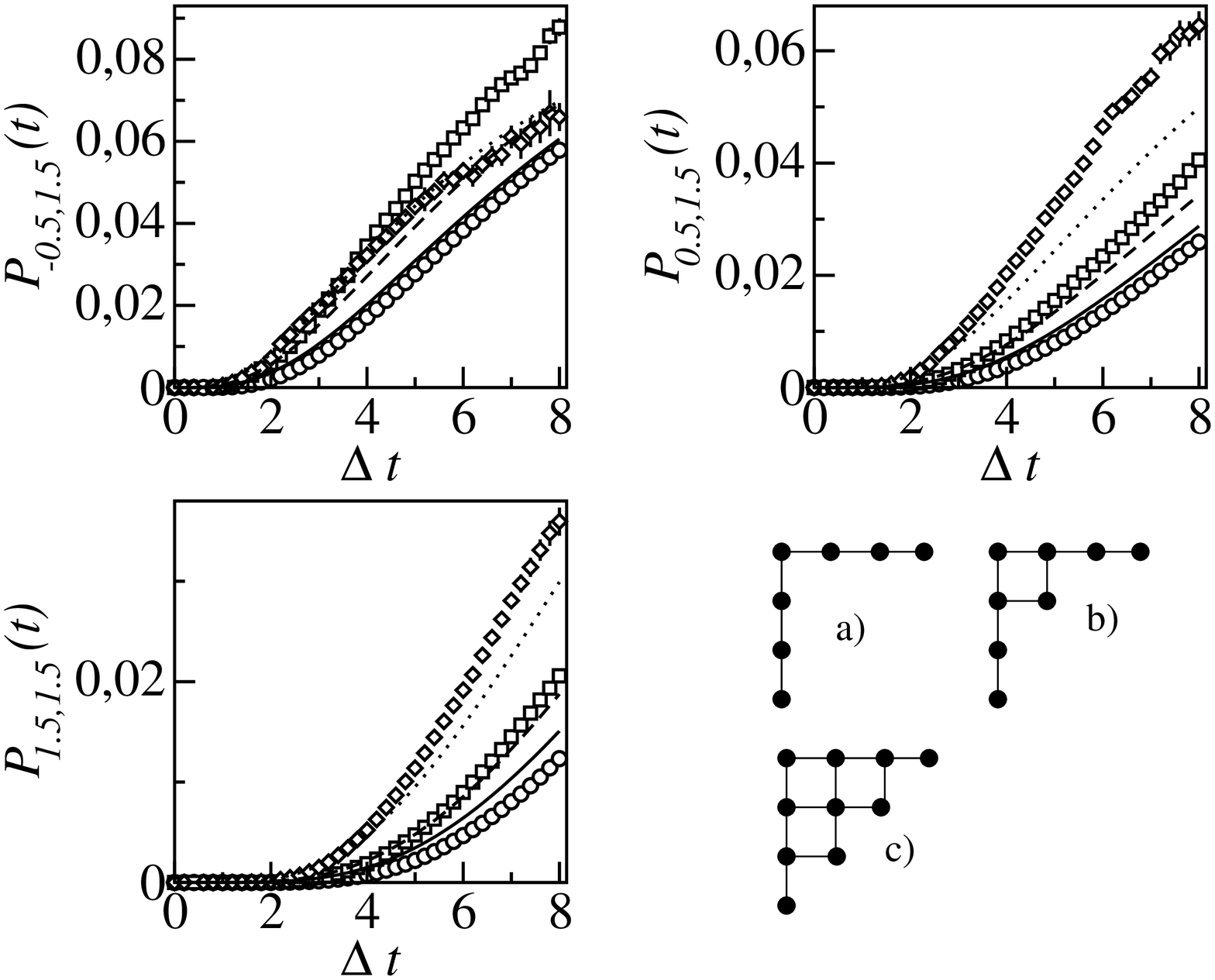,
width=8cm} \caption[]{\label{fig5} Dissipative dynamics of
two correlated charges in a one-dimensional tight binding
lattice with $d=5$ sites obtained from PIMC simulations
(symbols) and a master equation approach (lines). Shown
are the many-body state populations $P_{s,s'}$ for the
edge states of fig.~(\ref{fig4}) with either $s=-1.5$ or
$s'=1.5$. Circles, squares and diamonds denote situations
where some sites have been removed from otherwise
identical lattices as depicted in a), b), and c),
respectively, referring to an increasing connectivity
between the allowed sites. Solid (a), dashed (b), and
dotted lines (c), respectively, depict the corresponding
results from the master equation.}
\end{figure}

Let us now analyze (iii) in more detail based on
(\ref{differential rate equation}) with the corresponding
transition rates (\ref{grfull}). The simplest case is
$d=2$ with only three available sites
$(-\,{}^{1\!}/_2,-\,{}^{1\!}/_2),
(-\,{}^{1\!}/_2,\,{}^{1\!}/_2),
(\,{}^{1\!}/_2,\,{}^{1\!}/_2)$. Obviously, for this case
the triangular lattice coincides with a linear chain with
three sites so that if a time local master equation
applies for this latter case, it also applies for the
two-charge case. For larger $d$, however, the situation
changes fundamentally due to a different topology of the
two-dimensional lattices which then contain bulk-sites with
four adjacent sites (apart from $d=3$ where there is no
bulk site, but two edge-sites with three connections).
Accordingly, the typical dwell time on a bulk-site is
considerably shorter than on a site with two connections,
roughly, by a factor of 2. An environment which is able to
destroy the phase coherence of the wave function on each
site of a linear chain before a jump to an adjacent site
takes place, may be too slow to achieve the same on
bulk-sites in a 2d-lattice. Hence, entangled many body
states can survive on such a long time scale that a
description based on a master equation local in time
fails. In fact, this can be seen in fig.~\ref{fig5}, where
PIMC data for $d=4$ are depicted together with the
corresponding dynamics of the master equation
(\ref{differential rate equation}). By successively
removing more and more bulk sites from the lattice, the
agreement of the PIMC data with the prediction of the
master equation, which is rather poor in case of the full
lattice,
 increases, until upon creating a linear chain by
removing all bulk states an almost perfect match is
regained.

The situation becomes even worse for a larger number of charges $n$ since then
the bulk sites of the $n$-dimensional cube attain $2n$ decay channels to
adjacent sites, thus reducing the average dwell time by a factor of about $2n$
compared to the one-dimensional case $n=1$. The conclusion is the following: If
for a bath characterized by $\omega_c$ and $\beta$ a Markovian approximation
applies for the dynamics on a 1d-lattice with nearest neighbor coupling
$\Delta$, i.e.\ \ $\hbar\beta\omega_c <1$ and $\Delta/\omega_c\ll 1$, this
approximation fails for the multi-charge dynamics, unless $\omega_c$ is taken
to be very large and temperatures are sufficiently high, roughly, $n
\Delta/\omega_c\ll 1$ and $n \hbar\beta\omega_c <1$. Hence, for most cases only
numerical approaches like the PIMC procedure presented here, seem to be
applicable.

\section{Conclusions}

The numerically exact PIMC approach has been pushed further to deal also with
spatially continuous systems and correlated many-body dynamics. To better
understand the numerical data in certain parameter ranges, to estimate the
dissipative quantum dynamics before starting an involved PIMC calculation, and
even to develop simplified models capturing the relevant processes, Markovian
master equations are of great practical use. To explore their applicability in
the above situations, we had to consider generalized tight-binding lattices
with either non-equidistant spacing or in higher dimensions. In both cases,
additional restrictions must be imposed beyond the constraints known from
equidistant one-dimensional TBSs so that for broader ranges of bath parameters
no simple description seems to be possible. However, it is surprising that in
other domains comprising stronger dissipation and lower temperatures, Markovian
master equations can still be found to work quite well, at least
qualitatively. Thus, Markovian master equations together with numerically exact
PIMC simulations, which often provide the sole mean to check for their validity
in a certain scenario, provide a powerful means to study dynamical properties
over the full time range. In particular, for the correlated many-body dynamics
in presence of dissipation this may open the door to examine the intimate
interplay between Coulomb interaction, particle statistics, and phonon baths,
e.g.\ for the charge transport in molecular and mesoscopic structures.

\acknowledgements

This work benefited from fruitful discussions with A.
Komnik. We acknowledge financial support from the DFG
through Grant No.~AN336-1 and the Landesstiftung
Baden-W{\"u}rttemberg gGmbH. J.A.\ is a Heisenberg fellow
of the DFG.

\end{document}